\documentclass[10pt]{article}
\usepackage[cp866]{inputenc}
\begin{document}

\centerline {\Large\bf Qualitative investigation of Hamiltonian systems}
\centerline {\Large\bf by application of skew-symmetric differential forms}
\centerline {\large\bf L.I. Petrova}
\centerline{{\it Moscow State University, Russia, e-mail:ptr@cs.msu.su}}
\bigskip

A great number of works is devoted to qualitative investigation of
Hamiltonian systems. One of tools of such investigation is the
method of skew-symmetric differential forms [1-3].

In present work, under investigation Hamiltonian systems in
addition to skew-symmetric exterior differential forms,
skew-symmetric differential forms, which differ in their
properties from exterior forms, are used. These are skew-symmetric
differential forms defined on manifolds that are nondifferentiable
ones [4]. Such manifolds result, for example, under describing
physical processes by differential equations.

This approach to investigation of Hamiltonian systems enables one to
understand a connection between Hamiltonian systems and partial
differential equations, which describe physical processes, and to see
peculiarities of Hamiltonian systems and relevant phase spaces connected with
this fact.

\section{Connection between Hamiltonian systems and
partial differential equations}

The connection of Hamiltonian systems with partial differential
equations can be understood if one performs the analysis of partial
differential equations by means of skew-symmetric differential
forms. Such method of investigation has been developed by Cartan
[2] in his analysis of integrability of differential equations. In
present work we will call attention to some new aspects of such
investigation.

Let
$$ F(x^i,\,u,\,p_i)=0,\quad p_i\,=\,\partial u/\partial x^i \eqno(1)$$
be a partial differential equation of the first order.

Let us consider the functional relation
$$ du\,=\,\theta\eqno(2)$$
where $\theta\,=\,p_i\,dx^i$ (the summation over repeated indices is
implied). Here $\theta\,=\,p_i\,dx^i$ is a differential form of the
first degree.

The specific feature of functional relation (2) is that in the general
case, for example, when differential equation (1) describes any
physical processes, this relation turns out to be nonidentical.

The left-hand side of this relation involves a differential, and
the right-hand side includes the differential form
$\theta\,=\,p_i\,dx^i$. For this relation be identical, the
differential form $\theta\,=\,p_i\,dx^i$ must also  be a differential
(like the left-hand side of relation (2)), that is, it has to be a
closed exterior differential form. To do this, it requires the commutator 
$K_{ij}=\partial p_j/\partial x^i-\partial p_i/\partial x^j$ of the
differential form $\theta $ has to vanish.

In the general case, from equation (1) it does not follow (explicitly)
that the derivatives $p_i\,=\,\partial u/\partial x^i $, which obey
to the equation (and given boundary or initial conditions of the
problem), make up a differential. Without any supplementary conditions
the commutator $K_{ij}$ of the differential form $\theta $
is not equal to zero. The form $\theta\,=\,p_i\,dx^i$ turns out to be
unclosed and is not a differential like the left-hand side of relation
(2). Functional relation (2) appears to be nonidentical: the
left-hand side of this relation is a differential, whereas the right-hand
side is not a differential.

[Nonidentity of such relation has been pointed out in the work
[5]. In that case a possibility of using a symbol of differential
in the left-hand side of this relation has been allowed.] 

[Functional relation (2) can be written in the form 
$$ du\,-\,p_i\,dx^i=0\eqno(2')$$
This is a well-known Pfaff equation for partial differential equation. 
However, the relation cannot be treated as an equation. To solve the equation 
means to find the derivatives of equation (2') which make up a differential 
(and equation (2') is turned to identity). In this case the derivatives of 
equation (1) that do not obey these conditions are ignored, although 
they satisfy original equation (1) and boundary and initial conditions. 
But in the relation all derivatuves that satisfy the original equation and 
boundary or initial conditions are accounted for, and their role in the physical 
process under consideration is analyzed.]

The nonidentity of functional relation (2) points to a fact
that without additional conditions the derivatives of original
equation do not make up a differential. This means that the
corresponding solution $u$ of the differential equation will not be
a function of only variables $x^i$. The solution will depend on
the commutator of the form $\theta $, that is, it will be a functional.

To obtain a solution that is a function (i.e., the derivatives of
this solution compose a differential), it is necessary to add a
closure condition for the form $\theta\,=\,p_idx^i$ and for
corresponding dual form (in the present case the functional $F$
plays the role of a form dual to $\theta $) [2]:
$$\cases {dF(x^i,\,u,\,p_i)\,=\,0\cr
d(p_i\,dx^i)\,=\,0\cr}\eqno(3)$$ [The dual form that corresponds
to exterior differential form defines a manifold or structure, on
which the exterior form is defined.]

If we expand the differentials, we get a set of homogeneous equations
with respect to $dx^i$ and $dp_i$ (in the $2n$-dimensional tangent
space):
$$\cases {\displaystyle \left ({{\partial F}\over {\partial x^i}}\,+\,
{{\partial F}\over {\partial u}}\,p_i\right )\,dx^i\,+\,
{{\partial F}\over {\partial p_i}}\,dp_i \,=\,0\cr
dp_i\,dx^i\,-\,dx^i\,dp_i\,=\,0\cr} \eqno(4)$$

Solvability conditions for this system (vanishing of the determinant
composed of coefficients at $dx^i$, $dp_i$) have the form:
$$
{{dx^i}\over {\partial F/\partial p_i}}\,=\,{{-dp_i}\over {\partial F/\partial x^i+p_i\partial F/\partial u}} \eqno (5)
$$

The relations obtained establish a connection between the
differentials of coordinates $\{dx^i\}$ and differentials of
derivatives $\{dp_i\}$, which satisfy the original equation. It is
clear that these differentials specify integral curves on which
the derivatives of original equations form a differential. In
their properties the integral curves in phase space are
pseudostructures. The differential, which is defined only on
integral curve, is an interior differential. This differential
makes up a closed {\it inexact} exterior form, namely, an exterior
form closed only on some pseudostructure (to the pseudostructure
it is assigned the dual form).

Since on integral curves defined by relations (5) the derivatives of
equation (1) constitute a differential, the relevant solution to original equation
is a function rather then a functional. Such solutions that are functions (i.e.
depend only on variables) and are defined only on pseudostructures are
so-called generalized solutions [6].
Derivatives of generalized solution constitute
an exterior form, which is closed on the pseudostructure.

If conditions (5) are not satisfied, the differential form
$\theta\,=\,p_i\,dx^i$ is unclosed and is not a differential.
The derivatives do not form a differential, the solution that
corresponds to such derivatives will depend on the differential form
commutator $K_{ij}$ composed  of derivatives.
That means that the solution is a functional rather then a function.

Relations (5), which are integrating relations, can be obtained by
other means.
If we find the characteristics of equation (1), it turns out that
relations (5) are conditions which specify characteristics of the
equation under consideration. That is, integrating relations
(5) are characteristic relations for partial differential equation.
(In what follows such relations will be referred to as
characteristic relations).

Here it should call attention to some points that will be
necessary in carrying out further investigations.

Firstly, the characteristic relations have been obtained from a requirement of
vanishing the determinant (set of equations (4)). This means that a change from
original equation to the equation that obeys characteristic relations is
a degenerate transformation. One can see that this degenerate
transformation is a transition from derivatives of original equation in
tangent space to derivatives in cotangent space.

And secondly, to obtain the solution that is a function, it is necessary
to impose two (rather then one) additional conditions on the original equation;
1) The closure condition of the differential form composed of derivatives of
original equation, and 2) the condition of closure of relevant dual form.
The first condition is that the closed form is a differential. In this case,
as one can see from the results obtained, the closed form can be only an inexact
form, that is, this form is closed only on some pseudostructure. The second
condition just allows to obtain such pseudostricture.

\bigskip

Now assume that equation (1) does not depend explicitly on $u$
and is resolved with respect to some variable, for example $t$, that is,
the equation has the form
$$
{{\partial u}\over {\partial t}}\,+\,E(t,\,x^j,\,p_j)\,=\,0, \quad p_j\,=\,{{\partial u}\over {\partial x^j}}\eqno(6)
$$

In this case the differential form $\theta\,$ in functional
relation (2) will have the form $\theta\,=\,-Edt+\,p_j\,dx^j$

Relation (5) (the closure conditions of the
differential form $\theta $  and the corresponding dual form)
can be written as (in this case $\partial F/\partial p_1=1$)
$$
{{dx^j}\over {\partial E/\partial p_j}}\,=\,{{-dp_j}\over {\partial E/\partial x^j}}=dt
$$
and can be reduced to the form
$${{dx^j}\over {dt}}\,=\,{{\partial E}\over {\partial p_j}}, \quad
{{dp_j}\over {dt}}\,=\,-{{\partial E}\over {\partial x^j}}\eqno(7)$$

These are integral (characteristic) relations for equation (6), which
are conditions of integrability of this equation.

From relations (7) it follows that on integral curves the differential of
the form $\theta$ equals zero: $d\theta\,_{\pi}\,=\,d(-\,E\,dt\,+\,p_j\,dx^j)=0$
(here the index $\pi$ corresponds to integral curve, namely, to pseudostructure.
). This means that the derivatives of equation (6)
$p_1=\partial u/\partial t\,$, $p_j=\partial u/\partial x^j$
on integral curves obtained from relations (7) make up a closed
inexact exterior form $\theta\,_{\pi}=(-\,E\,dt+\,p_j\,dx^j)_{\pi}$, namely,
an interior, on integral curves, differential
$$(-\,E\,dt+\,p_j\,dx^j)_{\pi}\,=\,d_{\pi }\,u$$
The solution to equation (6) corresponding to such derivatives will be a function
(generalized) rather than a functional.

\bigskip

The equations of field theory have the form similar to that of equation (6)
$${{\partial s}\over {\partial t}}+H \left(t,\,q_j,\,p_j\right )\,=\,0,\quad
{{\partial s}\over {\partial q_j}}\,=\,p_j \eqno(8)$$
where $s$ is the field function (the state function) for the action
functional $S\,=\,\int\,L\,dt$.
Here $L(t,\,q_j,\,\dot q_j)$ is the Lagrange function, $H$ is the Hamilton function:
$H(t,\,q_j,\,p_j)\,=\,p_j\,\dot q_j-L$, $p_j\,=\,\partial L/\partial \dot q_j$.

Corresponding characteristic relations for equation (8) have the form
$${{dq_j}\over {dt}}\,=\,{{\partial H}\over {\partial p_j}}, \quad
{{dp_j}\over {dt}}\,=\,-{{\partial H}\over {\partial q_j}}\eqno(9)$$
that is, they are Hamiltonian systems.

As it is well known, the canonical relations have just such a form.

The analogy between Hamiltonian systems and characteristic relations,
as it will be shown below, allows one to see peculiarities of Hamiltonian
systems.

Here it should be emphasized that, in spite of equations (6) and
(8) have the same form, they fundamentally differ from one
another. As it is known, equation (8) is referred to as the
Hamilton-Jacobi equation. Unlike equation (6), where no
restrictions are imposed on the function $E$ and this function is
defined on tangent manifold, in equation (8) the function $H$ is
the Hamilton function defined on cotangent manifold, that is,
additional conditions are already imposed on this function. These
specific features will be considered below.

\section{Properties and peculiarities of characteristic relations}

As it has been shown above, the characteristic relations were obtained
from the first-order partial differential equation under the conditions that
the differential form composed of derivatives of this equation and
corresponding dual form have to be closed. Only under such conditions
the derivatives of original equation make up a differential, and corresponding
solutions prove to be functions rather than functionals, that is, they
depend only on variables.

Here it should be remembered that we analyze partial differential
equations which arise while describing physical processes. Without
additional conditions such equations are nonintegrable. The
characteristic relations are just additional conditions under
which the derivatives of original equation make up a differential.
They define integral curves $\{x^i(t), p_i(t)\}$ on which the
derivatives of original equation make up a closed inexact form,
namely, an interior differential.

What are peculiarities of characteristic relations?

We will analyze this by the example of relations (5).

As it was pointed above, characteristic relations (5) were obtained
from the condition that the determinant composed of coefficients at $dx^i$,
$dp_i$ of the set of equations (4) {\it vanish}.

This means that the characteristic relations are conditions of degenerate
transformation. (It turns out that only under vanishing some determinant, 
that is, under the condition of degenerate transformation, 
the derivatives of original equation can make up a differential.)

Peculiarities of characteristic relations (and, as it will be shown below,
of Hamiltonian systems) are just connected with the properties of degenerate
transformation.

The degenerate transformation can be mathematically presented as a transition
from one coordinate system to another, nonequivalent, one. In the case under
consideration this is a transition from tangent space, in which the derivatives
of original equation are defined, to cotangent space, in which the derivatives of
original equation make up a differential.

In the case under consideration the tangent space is not a differentiable 
manifold. (If the tangent space be differentiable, the differential 
of differential form $\theta\,=\,p_i\,dx^i$ would be equal to zero, that
is, this form would be closed one). The frame of reference
connected with such manifold cannot be an inertial system. In the
case of degeneration it takes place a transition from tangent
space to a manifold made up by pseudostructures (integral curves).
The frame of reference connected with such manifold is a locally
inertial one. That is, in this case the degenerate transformation
is a transition from the frame of reference which is not inertial
(and even cannot be locally inertial) to the locally inertial
frame of reference.

(It should be pointed out that, if the tangent manifold be differentiable,
the transition from tangent space to cotangent one would be not a degenerate
transformation. This would be a transition from one inertial frame of reference
to another inertial frame of reference.)

\bigskip
It turns out that the transition from derivatives of original equation,
which are defined in tangent space and do not make up a differential,
to derivatives that are defined in cotangent space and make up a differential,
is possible as a degenerate transformation. Since in this case cotangent
manifold and tangent manifold are not in one-to one correspondence, the
integral curves  it can serve only sections of cotangent bundle, 
namely, pseudostructures.

[Examples of pseudostructures and surfaces generated by them are
characteristics, cohomology, eikonal
surfaces, surfaces made up by shock wave fronts, potential surfaces, pseudo-Euclidean
and pseudo-Riemannian spaces and so on.]

Since the differential form composed of derivatives of original equation can
be closed only on pseudostructure, this form is an {\it inexact} exterior
differential form, that is, only an interior (on pseudostructure or on
integral curve) differential. And this means that corresponding solutions to original
equation, which are functions, are defined discretely, namely, only on pseudostructures.

As it was already pointed out, the solutions, which are defined on
pseudostructures and are functions, are so-called generalized solutions.
The derivatives of the generalized solution make up
the exterior form, which is closed on the pseudostructure.
(Under description of physical processes in material systems such solutions are
state functions because they have a differential.)

Since the functions, that are the generalized solutions, are defined
only on the pseudostructures (on integral curves), they have
{\it discontinuities in
derivatives in the directions normal to pseudostructures}.

To understand with what such discontinuities are connected and how much is their
value, one has to focus his attention to the following fact. The derivatives
of original equation simultaneously make up two skew-symmetric differential forms,
namely, one is an unclosed differential form composed of derivatives
of original equation and defined on tangent manifold, and the second
is a closed inexact exterior form defined on sections of cotangent bundles
(on pseudostructures).
The form closed on pseudostructure is an interior differential, and this enables one
to obtain the solution (generalized) to original equation on pseudostructure.
And the discontinuities
that have the derivatives of this solutions in the direction normal
to pseudostructure are specified by the commutator of the first-order unclosed
differential form. (Nonclosure of the first differential form is
connected with the fact that the tangent manifold corresponding to original equation
is not differentiable. In the case when tangent manifold is differentiable,
both first and second differential forms are closed and the derivatives of solutions
to original equation have no discontinuities. In should be noted that, for
differential equations describing physical processes, the tangent manifold
cannot be differentiable. For this reason the
solutions of all differential equations describing physical processes have
the above described functional properties[4]).

[The above considered functional properties of the set of
differential equations follow from kinematic and dynamical
conditions of consistency [7]. In the appendix to the work [8] the
results of calculating values of discontinuities of derivatives for 
entropy and sound speed in the gas dynamic problem are presented.]

\bigskip
Thus, it turns out that in general case the derivatives of partial
differential equations compose a differential only under degenerate
transformation. The characteristic relations are the conditions of such
degenerate transformation. The derivatives of differential equation obeying
characteristic relations make up an interior (on pseudostructure defined
by characteristic relation) differential and relevant solutions to original equation
are functions on pseudostructure. In this case the derivatives normal
to pseudostructure  undergo a discontinuity. Such specific features of characteristic
relations and the solutions corresponding to such relations enable one to see some
peculiarities of Hamiltonian systems and their relations to the equations of
mathematical physics.

\section{Analysis of Hamiltonian systems}

Hamiltonian systems arise in the problems of functional extremum
which have wide application in quantum field theory and in the
problems of classical mechanics at the basis of which it lie such
dynamic principles as the principle of minimal action, the
principle of virtual motions and so on.

Hamiltonian system (9) appears under the Legendre transformation:
$H(t,\,q_j,\,p_j)\,=\,p_j\,\dot q_j-L$,
$p_j\,=\,\partial L/\partial \dot q_j$,
which converts the Lagrange function $L(t,\, q_j,\dot q_j)$ defined on
tangent manifold $\{q_j,\dot q_j \}$ into the Hamilton function
$H(t,\,q_j,\,p_j)$ defined on cotangent manifold $\{q_j,p_j\}$.

The Hamiltonian system is connected with the Lagrange equation
$${{d}\over {dt}}\,{\,\partial L/\partial \dot q}\,-\,
{{\partial L}\over {\partial q}}\,=0 \eqno(10)$$
which specifies a curve that is an extremal of the functional.

The connection of Hamiltonian systems with the Lagrange equation can
be traced by comparing the differential of Hamilton function
$H(p,q,t)$ with the differential of the function $(p\,\dot q-L)$.
(Such comparison is presented in the work [1]. However, in present case
we shall focus our attention on some points of such comparison.)

The total differential of the Hamilton function $H(p,q,t)$ is written in the
form

$$dH=\,{{\partial H}\over {\partial p}}dp\,+
\,{{\partial H}\over {\partial q}}dq\,\,+
\,{{\partial H}\over {\partial t}}dt$$

And the total differential of the Hamilton function expressed in terms of
the Lagrange function  $H\,=\,p\,\dot q-L$ has the form
$$dH=\dot qdp\,-\,{{\partial L}\over {\partial q}}dq\,\,-
\,{{\partial L}\over {\partial t}}dt$$
These expressions will be identical under the condition
$$\dot q\,=\,{{\partial H}\over {\partial p}}, \quad
\,{{\partial L}\over {\partial q}}\,=-\,{{\partial H}\over {\partial q}}, \quad
\,{{\partial L}\over {\partial t}}\,=-\,{{\partial H}\over {\partial t}} \eqno(11)$$

From Lagrange equation (10) it follows that
${\,\partial L/\partial q}\,=\,\dot p$.
Replacing in the second relation (11) ${\,\partial L/\partial q}$
by $\dot p$, we obtain
$${{dp}\over {dt}}\,=\,-{{\partial H}\over {\partial q}}$$
which corresponds to the second relation of Hamiltonian system. That is, the second
relation of Hamiltonian system is just the Lagrange equation.

But from relations (11) one can see that under changing from the
Lagrange function to the Hamilton function in addition to the relation
corresponding to the Lagrange equation it arises one more relation, namely, the first
relation (11), which corresponds to the first relation for Hamiltonian system (9).
The physical meaning of such difference between Hamiltonian system and the Lagrange
equation will be analyzed below.

Thus, the connection of the Lagrange equation with Hamiltonian system is seen.
The transition from the Lagrange equation to Hamiltonian system is a transition
from tangent manifold to cotangent one. [Tangent and cotangent manifolds for
Lagrangian system are tangent and cotangent bundles of configuration space].
When tangent manifold is a differentiable one, such transition is
a degenerate transformation. The transition from tangent manifold to
cotangent one is one-to-one mapping, and Hamiltonian system and the Lagrange
equation are identical.

While deriving the Lagrange equation for mechanical system it was assumed that
constraints are ideal holonomic ones. In this case configuration space and
tangent manifold of Lagrangian system are differentiable manifolds [1], and
the transition from tangent space to cotangent one is a nondegenerate
transformation.

In the case on nonholonomic constraints the tangent manifold of Lagrangian
system will be not a differentiable manifold. In this case the transition
from tangent manifold to cotangent one, that is, the transition from the Lagrange function
to the Hamilton function and, correspondingly, from the Lagrange equation to Hamiltonian
system, is possible only as a degenerate transformation. This means that
the transition to subset of cotangent manifold composed of
pseudostructures (sections of cotangent bundles) is only possible. That is,
Hamiltonian system can be realized only discretely, namely, on pseudostructures.

In essence, Hamiltonian system will turn out to be a characteristic relation
for the Lagrange equation and will have the same peculiarities as
the characteristic relations.

In general case (when constrains are nonholonomic) the Lagrange equation is
nonintegrable equation. The solutions to the Lagrange equations define a curve
which is an extremal of the functional. But for these curves be integral curves,
the conditions of integrability have to be satisfied. The availability
of closed exterior form serves as the integrability condition.

The condition of maximum of the action functional $S$, from which
the Lagrange equation has been obtained, is one of conditions
being necessary in definition of closed form. But for the
differential form be closed, it is necessary that the relevant
dual form (determining manifold or structure on which the
skew-symmetric differential form is defined) be closed. The first
relation of Hamiltonian system is just such a condition. In the
case when tangent manifold of Lagrangian system is differentiable
(this is possible only for holonomic constrains), this condition
is satisfied automatically. Thus, Hamiltonian system is equivalent to the
Lagrange equation. In general case the
tangent manifold of Lagrangian system is not a differentiable
manifold, and hence the Lagrange equation can become integrable
one only under additional conditions. In this case the first relation
for Hamiltonian system proves to be such additional condition of
integrability of Lagrange equation. (In the calculus of variations
to such additional condition there corresponds the condition of
transversality.)

Thus, one can see that the Lagrange equation is equivalent to Hamiltonian system
only if the conditions of integrability are satisfied. In the case of nonholonomic
constrains when tangent manifold of Lagrangian system is not differentiable one,
such correspondence is satisfied only under degenerate transformation. The correspondence
between Hamiltonian system and the Lagrange equation is not identical.

What peculiarities in Hamiltonian system appear in the case when the
tangent manifold is nondifferentiable manifold and the transition
from Lagrange function to Hamilton function turns out to be degenerate
transformation?

As it was already pointed out,
under degenerate transformation the transition from tangent space is possible
only to pseudostructures. This means that as phase space formatted
it can be only the subset of cotangent manifold composed of pseudostructures
(sections of cotangent bundles). That is, in this case as the phase space it
can serve only cotangent bundle sections of manifold of Lagrangian system.

What properties has such phase space?

To answer this question, let us study a relation between Hamiltonian system and
the Hamilton-Jacobi equation.

It has been shown above the analogy between Hamiltonian system and
the characteristic relation for the first-order partial
differential equation. The Hamilton-Jacobi equation is an equation
of similar type. However, in the Hamilton-Jacobi equation it is
apriori assumed a fulfilment of additional conditions, namely, the
conditions of integrability. In this equation the function $H$ is
Hamilton function, that is, a function defined on cotangent and
not on tangent manifold. This fact points to a correspondence
between the Hamilton-Jacobi equation for state function and
Hamiltonian system.

Since the Hamiltonian system is fulfilled only on pseudostructures, 
the solutions to the Hamilton-Jacobi equation, which define the state 
function, can be only generalized functions. That is, the state function 
is defined only on pseudostructures, and the derivatives of state 
function have discontinuities in the direction normal to pseudostructure, 
namely, to the phase trajectory. Just with this fact the peculiarities of 
phase trajectories and phase space of Hamiltonian system are connected.

It is known that in the case when the tangent manifold is defferentiable
and hence when the transition from tangent space to cotangent space is 
one-to one mapping, in the extended phase space
$\{t,q_j,p_j\}$ there exists the Poincare invariant
$ds=\,-\,Hdt+\,p_jdx^j$ (the differential $ds$ directly follows
from the Hamilton-Jacobi equation). [Differential 
is invariant under gauge transformations (conserving the differential).
The canonical transformations are examples of such transformations.]

In the case when tangent manifold is not differentiable manifold
(and hence when the transition from tangent space to cotangent
space is degenerate) both Hamiltonian system and the
Hamilton-Jacobi equation will be fulfilled only on
pseudostructures, the Poincare invariant will be also fulfilled
only on pseudostructures, namely, on integral curves. In the
directions normal to integral curves the differential $ds$, which
corresponds to the Poincare invariant, will be discontinuous.

Invariants on pseudostructures set up invariant structures, which are
connected with conservation laws.

Closed exterior forms are invariants. The closed exterior form is
conservative quantity because the differential of closed form
equals zero. This means that the closed form reflects conservation
laws. And the closed {\it inexact} exterior form (a form closed on
pseudostructure) describes a conservative object, namely,
pseudostructure with conservative quantity. Such object is a
physical structure and corresponds to conservation law. Phase
trajectories with invariants (with closed forms) make up invariant
structures, which are physical structures corresponding to
conservation laws. Discontinuities (jumps) of invariants explain a
discreteness of physical structures. (It can be noted that such
invariant structures are an example of differential-geometric
G-structure).

As it was already pointed out above, Hamiltonian system is nothing more 
than canonical relations.

It is known that canonical relations execute {\it nondegenerate }
transformations, namely, transformations which conserve a
differential. The connection of Hamiltonian system with
characteristic and canonical relations discloses a duality of
Hamiltonian system. From one hand, in the case when tangent
manifold is not differentiable  manifold, Hamilton system
represents a characteristic relation, which is obtained as a
condition of degenerate transformation. And from another hand,
Hamiltonian system is canonical relations, which execute
nondegenerate transformation. The degenerate transformation is a
transition from {\it tangent} space ($q_j,\,\dot q_j)$) to {\it
cotangent} manifold ($q_j,\,p_j$). And the nondegenerate
transformation is a transition {\it in cotangent space} from some
pseudostructure  (phase trajectory) ($q_j,\,p_j$) to another
pseudostructure ($Q_j,\,P_j$). [The formula of canonical
transformation can be written as $p_jdq_j=P_jdQ_j+dW$, where $W$
is the generating function].

Thus, it turns out that Hamiltonian systems, from one hand (in the
case when the tangent manifold  $\{q_j,\,\dot q_j\}$ is not
differentiable one) are characteristic relations, which execute
degenerate transformations describing a transition from tangent
manifold on which there is no invariant structure, to cotangent
space, on which there is invariant structure. And from other hand,
Hamiltonian systems are canonical relations, which execute
nondegenerate transformations of invariant structures.

The transition from tangent space to cotangent one under
degenerate transformation when the closed exterior form is
realized describes an origination of invariant structure. And the
nondegenerate transformation (with the help of canonical
relations) is a transition from one invariant structures to
another invariant structure. (This demonstrates the connection of
degenerate and nondegenerate transformation.)

Nondegenerate transformations can be described by pseudogroups, in
particular, by Lie pseudogroups. But the group theory is not
sufficient for describing a behavior of Lagrangian systems in the
case of real physical processes.

1. Arnold V.~I. Mathematical methods of classical mechanics. -Moscow, 2003
(in Russian).

2. Cartan E., Les Systemes Differentials Exterieus ef Leurs Application
Geometriques. -Paris, Hermann, 1945.

3. Schutz B.~F., Geometric Methods of Mathematical Physics.
Cambridge University Press, Cambridge, 1982.

4. Petrova L.~I., Exterior and evolutionary skew-symmetric differential 
forms and their role in mathematical physics.

http://arXiv.org/pdf/math-ph/0310050

5. Synge J.~L. Tensorial Methods in Dynamics. -Department of
Applied Mathematics University of Toronto, 1936.

6. Vladimirov V.~S., Equations of mathematical physics. -Moscow,
Nauka, 1988 (in Russian).

7. Smirnov V.~I., A course of higher mathematics. -Moscow,
Tech.~Theor.~Lit. 1957, V.~4 (in Russian).

8. Petrova L.~I., The effect of noncommutativity of the conservation
laws on the development of thermodynamical and gas dynamical
instability.

http://arXiv.org/pdf/math-ph/0311040

\end{document}